\documentclass[12pt]{article}

\usepackage{amssymb,amsmath,amsthm,amscd,latexsym}
\usepackage{amsfonts}
\usepackage{tikz}
\usetikzlibrary{arrows,positioning,shapes,fit,calc}
\usepackage{tikz-cd}
\usepackage[mathscr]{eucal}
\usepackage{color,hyperref}
\usepackage{graphicx}
\usepackage{multirow}
\usepackage[ruled,vlined]{algorithm2e}
 \newtheorem{theorem}{Theorem}[section]

 \theoremstyle{definition}
 
 \theoremstyle{remark}

 \numberwithin{equation}{section}
\begin{document}

\title{New Type I Binary $[72,36,12]$ Self-Dual Codes from Composite Matrices and $R_1$ Lifts}
\author{ Adrian Korban \\
Department of Mathematical and Physical Sciences \\
University of Chester\\
Thornton Science Park, Pool Ln, Chester CH2 4NU, England \\
Serap \c{S}ahinkaya \\
Tarsus University, Faculty of Engineering \\ Department of Natural and Mathematical Sciences \\
Mersin, Turkey \\
Deniz Ustun \\
Tarsus University, Faculty of Engineering \\ Department of Computer Engineering \\
Mersin, Turkey}
 \maketitle

\begin{abstract}
In this work, we define three composite matrices derived from group rings. We employ these composite matrices to create generator matrices of the form $[I_n \ | \ \Omega(v)],$ where $I_n$ is the identity matrix and $\Omega(v)$ is a composite matrix and search for binary self-dual codes with parameters $[36,18, 6 \ \text{or} \ 8].$ We next lift these codes over the ring $R_1=\mathbb{F}_2+u\mathbb{F}_2$ to obtain codes whose binary images are self-dual codes with parameters $[72,36,12].$ Many of these codes turn out to have weight enumerators with parameters that were not known in the literature before. In particular, we find $30$ new Type I binary self-dual codes with parameters $[72,36,12].$ 
\end{abstract}

\section{Introduction}\label{intro}

In this work, we employ the idea of composite matrices from group rings that was introduced in \cite{DoughertyIII}, to search for binary self-dual codes. In particular, we consider a generator matrix of the form $[I_n \ | \ \Omega(v)],$ where $I_n$ is the identity matrix and $\Omega(v)$ is a composite matrix where $v$ is an element in the group ring $RG,$ to search for binary self-dual codes. Similar approach for finding extremal binary self-dual codes of length 68 can be found in \cite{Korban2, Korban3}. We describe three composite matrices from group rings with the use of groups of orders $18, 9$ and $6$ which we then employ to search for binary self-dual codes with parameters $[36,18, 6 \ \text{or} \ 8].$ We next lift these codes over the ring $\mathbb{F}_2+u\mathbb{F}_2,$ to obtain codes whose binary images are self-dual codes of length 72. Many of these codes turn out to be new Type I binary self-dual codes with parameters $[72,36,12].$ 

The rest of the work is organized as follows. In Section~2, we give preliminary definitions and results on codes, the alphabets we use, special matrices, group rings and composite matrices from group rings. In Section 3, we construct three composite matrices of order $18$ by using group rings and groups of different cardinalities, in particular, groups with cardinalities $18, 9$ and $6.$ Next, in Section~4, we define three generator matrices in which we use the composite matrices described in Section~3. We show the conditions that each generator matrix needs to meet in order to produce self-dual codes. In Section~5, we present all the computational results obtained from searching for self-dual codes with different parameters and over different alphabets and with the use of the generator matrices presented in Section~4. We finish with concluding remarks and directions for possible future research. 

\section{Preliminaries}

\subsection{Codes}

We begin by recalling the standard definitions from coding theory. A code $C$
of length $n$ over a Frobenius ring $R$ is a subset of $R^n$. If the code is
a submodule of $R^n$ then we say that the code is linear. Elements of the
code $C$ are called codewords of $C$. Let $\mathbf{x}=(x_1,x_2,\dots,x_n)$
and $\mathbf{y}=(y_1,y_2,\dots,y_n)$ be two elements of $R^n.$ The duality
is understood in terms of the Euclidean inner product, namely:
\begin{equation*}
\langle \mathbf{x},\mathbf{y} \rangle_E=\sum x_iy_i.
\end{equation*}
The dual $C^{\bot}$ of the code $C$ is defined as
\begin{equation*}
C^{\bot}=\{\mathbf{x} \in R^n \ | \ \langle \mathbf{x},\mathbf{y}
\rangle_E=0 \ \text{for all} \ \mathbf{y} \in C\}.
\end{equation*}
We say that $C$ is self-orthogonal if $C \subseteq C^\perp$ and is self-dual
if $C=C^{\bot}.$

An upper bound on the minimum Hamming distance of a binary self-dual code
was given in \cite{RainsI}. Specifically, let $d_{I}(n)$ and $d_{II}(n)$ be the
minimum distance of a Type~I and Type~II binary code of length $n$,
respectively. Then
\begin{equation*}
d_{II}(n) \leq 4\lfloor \frac{n}{24} \rfloor+4
\end{equation*}
and
\begin{equation*}
d_{I}(n)\leq
\begin{cases}
\begin{matrix}
4\lfloor \frac{n}{24} \rfloor+4 \ \ \ if \ n \not\equiv 22 \pmod{24} \\
4\lfloor \frac{n}{24} \rfloor+6 \ \ \ if \ n \equiv 22 \pmod{24}.%
\end{matrix}%
\end{cases}%
\end{equation*}

Self-dual codes meeting these bounds are called \textsl{extremal}.
Throughout the text, we obtain extremal binary codes of different lengths.
Self-dual codes which are the best possible for a given set of parameters is
said to be optimal. Extremal codes are necessarily optimal but optimal codes
are not necessarily extremal.

\subsection{The ring $R_1=\mathbb{F}_2+u\mathbb{F}_2$}

In this section, we recall some theory on self-dual codes over $\mathbb{F}%
_2+u\mathbb{F}_2.$ We refer to \cite{DoughertyI} where Type II, Type IV, self-dual
codes and cyclic codes over $\mathbb{F}_2+u\mathbb{F}_2$ were studied.

The ring $\mathbb{F}_2+u\mathbb{F}_2$ is a ring of characteristic 2 with 4
elements with the restriction $u^2=0.$ It is defined as
\begin{equation*}
\mathbb{F}_2+u\mathbb{F}_2=\{a+bu \ | \ a,b \in \mathbb{F}_2, u^2=0\},
\end{equation*}
and it is easily seen that $\mathbb{F}_2+u\mathbb{F}_2 \cong \mathbb{F}%
_2[x]/(x^2).$ A linear code $C$ of length $n$ over the ring $\mathbb{F}_2+u%
\mathbb{F}_2$ is an $\mathbb{F}_2+u\mathbb{F}_2$-submodule of $(\mathbb{F}%
_2+u\mathbb{F}_2)^n.$ The elements of $\mathbb{F}_2+u\mathbb{F}_2$ are $%
0,1,u,1+u$ and their Lee weights are defined as $0,1,2,1$ respectively. The
Hamming $(d_H)$ and Lee $(d_L)$ distance between $n$ tuples is then defined
as the sum of the Hamming and Lee weights of the difference of the
components of these tuples respectively. The smallest positive Hamming and
Lee distance of a code $C$ is denoted by $d_H(C)$ and $d_L(C)$ respectively.

A Gray map $\phi$ is defined as
\begin{equation*}
\phi : (\mathbb{F}_2+u\mathbb{F}_2)^n \rightarrow \mathbb{F}_2^{2n},
\end{equation*}
\begin{equation*}
\phi(\overline{a}+\overline{b}u)=(\overline{b},\overline{a}+\overline{b}),
\end{equation*}
where $\overline{a}, \overline{b} \in \mathbb{F}_2^n.$ The map is a distance
preserving isometry from $((\mathbb{F}_2+u\mathbb{F}_2)^n, d_L)$ to $(%
\mathbb{F}_2^{2n},d_H),$ where $d_L$ and $d_H$ denote the Lee and Hamming
distance in $(\mathbb{F}_2+u\mathbb{F}_2)^n$ and $\mathbb{F}_2^{2n}$
respectively. This means that if $C$ is a linear code over $\mathbb{F}_2+u%
\mathbb{F}_2$ with parameters $[n,2^k,d]$ ($2^k$ is the number of the
codewords), then $\phi(C)$ is a binary linear code of parameters $[2n,k,d].$
The following theorem is a natural result of the Gray map.

\begin{theorem}
\label{thm.2.2.1} If $C$ is a self-dual code over $\mathbb{F}_2+u\mathbb{F}_2
$ of length $n,$ then $\phi(C)$ is a self-dual binary code of length $2n.$
\end{theorem}

We can also define a natural projection from $\mathbb{F}_2+u\mathbb{F}_2$ to
$\mathbb{F}_2$ as follows:
\begin{equation*}
\mu : \mathbb{F}_2+u\mathbb{F}_2 \rightarrow \mathbb{F}_2,
\end{equation*}
\begin{equation*}
\mu(a+bu)=a.
\end{equation*}
If $D=\mu(C)$ for some linear code $C$ over $\mathbb{F}_2+u\mathbb{F}_2,$ we
say that $D$ is a projection of $C$ into $\mathbb{F}_2,$ and that $C$ is a
lift of $D$ into $\mathbb{F}_2+u\mathbb{F}_2.$ It is clear that the
projection of a self-orthogonal code is self-orthogonal, but the projection
of a self-dual code need not be self-dual. We finish this section with two
well known results.

\begin{theorem}
\label{thm.2.2.2} Suppose that $C$ is a self-dual code over $\mathbb{F}_2+u%
\mathbb{F}_2$ of length $2n,$ generated by the matrix $[I_n|A],$ where $I_n$
is the $n \times n$ identity matrix. Then $\mu(C)$ is a self-dual binary
code of length $2n.$
\end{theorem}

\begin{theorem}
\label{thm.2.2.3} Suppose $C$ is a linear code over $\mathbb{F}_2+u\mathbb{F}%
_2$ and that $C^{\prime }=\mu(C),$ is its projection to $\mathbb{F}_2.$ With
$d$ and $d^{\prime }$ representing the minimum Lee and Hamming distances of $%
C$ and $C^{\prime }$ respectively, we have that $d \leq 2d^{\prime }.$
\end{theorem}

\subsection{Special Matrices and Group Rings}

To understand the composite construction which we define later in this work,
we recall some basic definitions of some special matrices and theory on group rings and the map that
sends group ring elements to matrices.

A circulant matrix is one where each row is shifted one element to the right relative to the preceding row. We label the circulant matrix as $A=circ(\alpha_1,\alpha_2\dots , \alpha_n),$ where $\alpha_i$ are ring elements. The transpose of a matrix $A,$ denoted by $A^T,$ is a matrix whose rows are the columns of $A,$ i.e., $A^T_{ij}=A_{ji}.$  A symmetric matrix is a square matrix that is equal to its transpose.

While group rings can be given for infinite rings and infinite groups, we
are only concerned with group rings where both the ring and the group are
finite. Let $G$ be a finite group of order $n$, then the group ring $RG$
consists of $\sum_{i=1}^n \alpha_i g_i$, $\alpha_i \in R$, $g_i \in G.$

Addition in the group ring is done by coordinate addition, namely
\begin{equation}
\sum_{i=1}^n \alpha_i g_i +\sum_{i=1}^n \beta_i g_i =\sum_{i=1}^n (\alpha_i
+ \beta_i)g_i.
\end{equation}
The product of two elements in a group ring is given by
\begin{equation}
\left(\sum_{i=1}^n \alpha_i g_i \right)\left(\sum_{j=1}^n \beta_j g_j
\right)= \sum_{i,j} \alpha_i \beta_j g_i g_j.
\end{equation}
It follows that the coefficient of $g_k$ in the product is $\sum_{g_i
g_j=g_k} \alpha_i \beta_j.$

The following construction of a matrix was first given for codes over fields
by Hurley in \cite{HurleyI}. It was extended to Frobenius rings in \cite{DoughertyII}. Let $%
R$ be a finite commutative Frobenius ring and let $G=\{g_1,g_2,\dots,g_n\}$
be a group of order $n$ and let $v=\sum_{i=1}^n \alpha_{g_i} \in RG.$ Define
the matrix $\sigma(v) \in M_n(R)$ to be 
\begin{equation}\label{sigmav}
\sigma(v)=%
\begin{pmatrix}
\alpha_{g_1^{-1}g_1} & \alpha_{g_1^{-1}g_2} & \alpha_{g_1^{-1}g_3} & \dots &
\alpha_{g_1^{-1}g_n} \\
\alpha_{g_2^{-1}g_1} & \alpha_{g_2^{-1}g_2} & \alpha_{g_2^{-1}g_3} & \dots &
\alpha_{g_2^{-1}g_n} \\
\vdots & \vdots & \vdots & \vdots & \vdots \\
\alpha_{g_n^{-1}g_1} & \alpha_{g_n^{-1}g_2} & \alpha_{g_n^{-1}g_3} & \dots &
\alpha_{g_n^{-1}g_n}%
\end{pmatrix}%
.
\end{equation} We note that the elements $g_1^{-1},
g_2^{-1}, \dots, g_n^{-1}$ are the elements of the group $G$ in a some given
order.

\subsection{Composite Matrices from Group Rings}

In this section, we recall a matrix construction which is an extension of the matrix $\sigma(v)$ in Equation~(\ref{sigmav}). This extension was first introduced in \cite{DoughertyIII}. With this extension, one can produce many interesting $n \times n$ matrices for different choices of not just one group as in the case of the matrix $\sigma(v),$ but for different choices of more than one group, please see \cite{DoughertyIII} for more details. Such matrices are referred to as composite constructions or just composite matrices.

Let $R$ be a finite commutative Frobenius ring. Let $\{g_1,g_2,\dots ,g_n\}$ be a fixed listing of the elements of $G.$ Let $\{(h_i)_1,(h_i)_2,\dots ,(h_i)_r\}$  be a fixed listing of the elements of $H_i,$ where $H_i$ is any group of order $r.$ Let $r$ be a factor of $n$ with $n>r$ and $n,r \neq 1.$ Also, let $G_r$ be a subset of $G$ containing $r$ distinct elements of $G.$ Define the map:

\begin{table}[h!]
\centering
\begin{tabular}{ccc}
\multicolumn{3}{c}{$\phi: H \mapsto G_r$}                \\
$h_1$ & $\xrightarrow{\phi}$ & $g_1$         \\
$h_2$ & $\xrightarrow{\phi}$ & $g_2$     \\
$\vdots$  & $\vdots$             & $\vdots$                \\
$h_r$ & $\xrightarrow{\phi}$ & $g_r.$
\end{tabular}
\end{table}

This map sends $r$ distinct elements of the group $H$ to $r$ distinct elements of the group $G.$

Let $v=\alpha_{g_1}g_1+\alpha_{g_2}g_2+\dots,\alpha_{g_n}g_n \in RG.$ Define the matrix $\Omega(v) \in M_n(R)$ to be

\begin{equation}\label{Composite construction}
\Omega(v)=\begin{pmatrix}
A_1&A_2&A_3&\dots &A_{\frac{n}{r}}\\
A_{\frac{n}{r}+1}&A_{\frac{n}{r}+2}&A_{\frac{n}{r}+3}&\dots &A_{\frac{2n}{r}}\\
\vdots & \vdots & \vdots & \vdots & \vdots \\
A_{\frac{(r-1)n}{r}+1}&A_{\frac{(r-1)n}{r}+2}&A_{\frac{(r-1)n}{r}+3}&\dots & A_{\frac{n^2}{r^2}}
\end{pmatrix},
\end{equation}
where at least one block has the following form:
$$A_l=\begin{pmatrix}
\alpha_{g_j^{-1}g_k}&\alpha_{g_j^{-1}g_{k+1}}&\dots &\alpha_{g_j^{-1}g_{k+(r-1)}}\\
\alpha_{g_{j+1}^{-1}g_k}&\alpha_{g_{j+1}^{-1}g_{k+1}}&\dots &\alpha_{g_{j+1}^{-1}g_{k+(r-1)}}\\
\alpha_{g_{j+2}^{-1}g_k}&\alpha_{g_{j+2}^{-1}g_{k+1}}&\dots &\alpha_{g_{j+2}^{-1}g_{k+(r-1)}}\\
\vdots & \vdots & \vdots & \vdots\\
\alpha_{g_{j+r-1}^{-1}g_k}&\alpha_{g_{j+r-1}^{-1}g_{k+1}}&\dots &\alpha_{g_{j+r-1}^{-1}g_{k+(r-1)}}
\end{pmatrix},$$
and the other blocks are of the form:
$$A_l'=\begin{pmatrix}
\alpha_{g_j^{-1}g_k}&\alpha_{g_j^{-1}g_{k+1}}&\dots &\alpha_{g_j^{-1}g_{k+(r-1)}}\\
\alpha_{\phi_l((h_i)_2^{-1}(h_i)_1)}&\alpha_{\phi_l((h_i)_2^{-1}(h_i)_2)}&\dots &\alpha_{\phi_l((h_i)_2^{-1}(h_i)_r)}\\
\alpha_{\phi_l((h_i)_3^{-1}(h_i)_1)}&\alpha_{\phi_l((h_i)_3^{-1}(h_i)_2)}&\dots &\alpha_{\phi_l((h_i)_3^{-1}(h_i)_r)}\\
\vdots & \vdots & \vdots & \vdots\\
\alpha_{\phi_l((h_i)_r^{-1}(h_i)_1)}&\alpha_{\phi_l((h_i)_r^{-1}(h_i)_2)}&\dots &\alpha_{\phi_l((h_i)_r^{-1}(h_i)_r)}\\
\end{pmatrix},$$
where $l=\{1,2,3,\dots ,\frac{n^2}{r^2}\}$ and where:

\begin{table}[h!]
\centering
\begin{tabular}{ccc}
\multicolumn{3}{c}{$\phi_l: H_i \mapsto G_r$}                \\
$(h_i)_1$ & $\xrightarrow{\phi_l}$ & $g_{j}^{-1}g_k$         \\
$(h_i)_2$ & $\xrightarrow{\phi_l}$ & $g_{j}^{-1}g_{k+1}$     \\
$\vdots$  & $\vdots$             & $\vdots$                \\
$(h_i)_r$ & $\xrightarrow{\phi_l}$ & $g_{j}^{-1}g_{k+(r-1)}.$
\end{tabular}
\end{table}

Here we notice that when when $l=1$ then $j=1,k=1,$ when $l=2$ then $j=1,k=r+1,$ when $l=3$ then $j=1,k=2r+1,$ $\dots$ when $l=\frac{n}{r}$ then $j=1,k=n-r+1.$ When $l=\frac{n}{r}+1$ then $j=r+1, k=1,$ when $l=\frac{n}{r}+2$ then $j=r+1, k=r+1,$ when $l=\frac{n}{r}+3$ then $j=r+1, k=2r+1,$ $\dots$ when $l=\frac{2n}{r}$ then $j=r+1, k=n-r+1,$ $\dots,$ and so on.

\section{Composite Matrices}

In this section, we employ the matrix construction given in Equation~(\ref{Composite construction}) to build three composite matrices. We particularly consider some groups of orders $18, 9$ and $6$ to construct composite matrices of order $18$ which we later use to search for binary self-dual codes.

\begin{enumerate}
\item[1.] Let $G=\langle x,y \ | \ x^9=y^2=x^y=x^{-1} \rangle \cong D_{18}.$ Let $v_1=\sum_{i=0}^8 \sum_{j=0}^1 \alpha_{1+i+9j}y^jx^i \in RD_{18}.$ Also, let $H=\langle a \ | \ a^{9}=1 \rangle \cong C_{3,3}.$ We now define the composite matrix over $R$ as
$$\Omega(v_1)=\begin{pmatrix}
A_1'&A_2'\\
A_3'&A_4'
\end{pmatrix}$$
where
$$A_1'=\begin{pmatrix}
\alpha_{g_1^{-1}g_1}&\alpha_{g_1^{-1}g_2}&\alpha_{g_1^{-1}g_3}& \dots &\alpha_{g_1^{-1}g_9}\\
\alpha_{\phi_1(h_2^{-1}h_1)}&\alpha_{\phi_1(h_2^{-1}h_2)}&\alpha_{\phi_1(h_2^{-1}h_3)}& \dots &\alpha_{\phi_1(h_2^{-1}h_9)}\\

\alpha_{\phi_1(h_3^{-1}h_1)}&\alpha_{\phi_1(h_3^{-1}h_2)}&\alpha_{\phi_1(h_3^{-1}h_3)}& \dots &\alpha_{\phi_1(h_3^{-1}h_9)}\\

\vdots&\vdots&\vdots&\vdots&\vdots\\

\alpha_{\phi_1(h_9^{-1}h_1)}&\alpha_{\phi_1(h_9^{-1}h_2)}&\alpha_{\phi_1(h_9^{-1}h_3)}& \dots &\alpha_{\phi_1(h_9^{-1}h_9)}\\

\end{pmatrix}$$
with
$$\phi_1 : h_j^{-1}h_i \xrightarrow{\phi_1} g_1^{-1}g_i$$
$$ \text{for when} $$
$$ j=2: \ i=1,2,\dots,9$$
$$ j=3: \ i=1,2,\dots,9$$
$$ \vdots $$
$$ j=9: \ i=1,2,\dots,9,$$

$$A_2'=\begin{pmatrix}
\alpha_{g_1^{-1}g_{10}}&\alpha_{g_1^{-1}g_{11}}&\alpha_{g_1^{-1}g_{12}}& \dots &\alpha_{g_1^{-1}g_{18}}\\
\alpha_{\phi_2(h_2^{-1}h_1)}&\alpha_{\phi_2(h_2^{-1}h_2)}&\alpha_{\phi_2(h_2^{-1}h_3)}& \dots &\alpha_{\phi_2(h_2^{-1}h_9)}\\

\alpha_{\phi_2(h_3^{-1}h_1)}&\alpha_{\phi_2(h_3^{-1}h_2)}&\alpha_{\phi_2(h_3^{-1}h_3)}& \dots &\alpha_{\phi_2(h_3^{-1}h_9)}\\

\vdots&\vdots&\vdots&\vdots&\vdots\\

\alpha_{\phi_2(h_9^{-1}h_1)}&\alpha_{\phi_2(h_9^{-1}h_2)}&\alpha_{\phi_2(h_9^{-1}h_3)}& \dots &\alpha_{\phi_2(h_9^{-1}h_9)}\\

\end{pmatrix}$$
with
$$\phi_2 : h_j^{-1}h_k \xrightarrow{\phi_2} g_1^{-1}g_i$$
$$ \text{for when} $$
$$ j=2: \ k=1, i=10; \ k=2, i=11; \ \dots \ k=9, i=18,$$
$$ j=3: \ k=1, i=10; \ k=2, i=11; \ \dots \ k=9, i=18,$$
$$ \vdots $$
$$ j=9: \ k=1, i=10; \ k=2, i=11; \ \dots \ k=9, i=18,$$

$$A_3'=\begin{pmatrix}
\alpha_{g_{10}^{-1}g_1}&\alpha_{g_{10}^{-1}g_2}&\alpha_{g_{10}^{-1}g_3}& \dots &\alpha_{g_{10}^{-1}g_9}\\
\alpha_{\phi_3(h_2^{-1}h_1)}&\alpha_{\phi_3(h_2^{-1}h_2)}&\alpha_{\phi_3(h_2^{-1}h_3)}& \dots &\alpha_{\phi_3(h_2^{-1}h_9)}\\

\alpha_{\phi_3(h_3^{-1}h_1)}&\alpha_{\phi_3(h_3^{-1}h_2)}&\alpha_{\phi_3(h_3^{-1}h_3)}& \dots &\alpha_{\phi_3(h_3^{-1}h_9)}\\

\vdots&\vdots&\vdots&\vdots&\vdots\\

\alpha_{\phi_3(h_9^{-1}h_1)}&\alpha_{\phi_3(h_9^{-1}h_2)}&\alpha_{\phi_3(h_9^{-1}h_3)}& \dots &\alpha_{\phi_3(h_9^{-1}h_9)}\\

\end{pmatrix}$$
with
$$\phi_3 : h_j^{-1}h_i \xrightarrow{\phi_3} g_{10}^{-1}g_i$$
$$ \text{for when} $$
$$ j=2: \ i=1,2,\dots,9$$
$$ j=3: \ i=1,2,\dots,9$$
$$ \vdots $$
$$ j=9: \ i=1,2,\dots,9,$$
and

$$A_4'=\begin{pmatrix}
\alpha_{g_{10}^{-1}g_{10}}&\alpha_{g_{10}^{-1}g_{11}}&\alpha_{g_{10}^{-1}g_{12}}& \dots &\alpha_{g_{10}^{-1}g_{18}}\\
\alpha_{\phi_4(h_2^{-1}h_1)}&\alpha_{\phi_4(h_2^{-1}h_2)}&\alpha_{\phi_4(h_2^{-1}h_3)}& \dots &\alpha_{\phi_4(h_2^{-1}h_9)}\\

\alpha_{\phi_4(h_3^{-1}h_1)}&\alpha_{\phi_4(h_3^{-1}h_2)}&\alpha_{\phi_4(h_3^{-1}h_3)}& \dots &\alpha_{\phi_4(h_3^{-1}h_9)}\\

\vdots&\vdots&\vdots&\vdots&\vdots\\

\alpha_{\phi_4(h_9^{-1}h_1)}&\alpha_{\phi_4(h_9^{-1}h_2)}&\alpha_{\phi_4(h_9^{-1}h_3)}& \dots &\alpha_{\phi_4(h_9^{-1}h_9)}\\

\end{pmatrix}$$
with
$$\phi_4 : h_j^{-1}h_k \xrightarrow{\phi_4} g_{10}^{-1}g_i$$
$$ \text{for when} $$
$$ j=2: \ k=1, i=10; \ k=2, i=11; \ \dots \ k=9, i=18,$$
$$ j=3: \ k=1, i=10; \ k=2, i=11; \ \dots \ k=9, i=18,$$
$$ \vdots $$
$$ j=9: \ k=1, i=10; \ k=2, i=11; \ \dots \ k=9, i=18.$$
This results in a composite matrix over $R$ of the following form:
$$\Omega(v_1)=\begin{pmatrix}
A_1'&A_2'\\
A_3'&A_4'
\end{pmatrix}=\begin{pmatrix}
B&C\\
D&E
\end{pmatrix}=$$
$$=\begin{pmatrix}
B_1&B_2&B_3&C_1&C_2&C_3\\
B_3'&B_1&B_2&C_3'&C_1&C_2\\
B_2'&B_3'&B_1&C_2'&C_3'&C_1\\
D_1&D_2&D_3&E_1&E_2&E_3\\
D_3'&D_1&D_2&E_3'&E_1&E_2\\
D_2'&D_3'&D_1&E_2'&E_3'&E_1
\end{pmatrix},$$
where 
$$B_1=circ(\alpha_1,\alpha_2,\alpha_3), B_2=circ(\alpha_4,\alpha_5,\alpha_6), B_3=circ(\alpha_7,\alpha_8,\alpha_9),$$
$$B_2'=circ(\alpha_6,\alpha_4,\alpha_5), B_3'=circ(\alpha_9,\alpha_7,\alpha_8), C_1=circ(\alpha_{10},\alpha_{11},\alpha_{12}),$$
$$C_2=circ(\alpha_{13},\alpha_{14},\alpha_{15}), C_3=circ(\alpha_{16},\alpha_{17},\alpha_{18}), C_2'=circ(\alpha_{15},\alpha_{13},\alpha_{14}),$$
$$C_3'=circ(\alpha_{18},\alpha_{16},\alpha_{17}), D_1=circ(\alpha_{10},\alpha_{18},\alpha_{17}), D_2=circ(\alpha_{16},\alpha_{15},\alpha_{14}),$$
$$D_3=circ(\alpha_{13},\alpha_{12},\alpha_{11}), D_2'=circ(\alpha_{14},\alpha_{16},\alpha_{15}), D_3'=circ(\alpha_{11},\alpha_{13},\alpha_{12}),$$
$$E_1=circ(\alpha_1,\alpha_9,\alpha_8), E_2=circ(\alpha_7,\alpha_6,\alpha_4), E_3=circ(\alpha_3,\alpha_2,\alpha_1),$$
$$E_2'=circ(\alpha_4,\alpha_7,\alpha_6) \ \text{and} \ E_3'=circ(\alpha_1,\alpha_3,\alpha_2).$$

\item[2.] Let $G=\langle x,y \ | \ x^6=y^3=1, xy=yx \rangle \cong C_3 \times C_6.$ Let $v_2=\sum_{i=0}^5 \sum_{j=0}^2 \alpha_{1+i+6j}x^iy^j \in R(C_3 \times C_6).$ Also, let $H=\langle a,b \ | \ a^3=b^2=1, a^b=a^{-1} \rangle \cong D_6.$ We now define the composite matrix over $R$ as
$$\Omega(v_2)=\begin{pmatrix}
A_1'&A_2'&A_3'\\
A_4'&A_5'&A_6'\\
A_7'&A_8'&A_9'
\end{pmatrix}$$
where
$$A_1'=\begin{pmatrix}
\alpha_{g_1^{-1}g_1}&\alpha_{g_1^{-1}g_2}&\dots &\alpha_{g_1^{-1}g_6}\\
\alpha_{\phi_1(h_2^{-1}h_1)}&\alpha_{\phi_1(h_2^{-1}h_2)}&\dots &\alpha_{\phi_1(h_2^{-1}h_6)}\\
\alpha_{\phi_1(h_3^{-1}h_1)}&\alpha_{\phi_1(h_3^{-1}h_2)}&\dots &\alpha_{\phi_1(h_3^{-1}h_6)}\\
\vdots & \vdots & \vdots & \vdots \\
\alpha_{\phi_1(h_6^{-1}h_1)}&\alpha_{\phi_1(h_6^{-1}h_2)}&\dots &\alpha_{\phi_1(h_6^{-1}h_6)}\\
\end{pmatrix}$$
with 
$$\phi_1 : h_j^{-1}h_i \xrightarrow{\phi_1} g_1^{-1}g_i$$
$$ \text{for when} $$
$$ j=2: \ i=1,2,\dots,6$$
$$ j=3: \ i=1,2,\dots,6$$
$$ \vdots $$
$$ j=6: \ i=1,2,\dots,6,$$

$$A_2'=\begin{pmatrix}
\alpha_{g_1^{-1}g_7}&\alpha_{g_1^{-1}g_8}&\dots &\alpha_{g_1^{-1}g_{12}}\\
\alpha_{\phi_2(h_2^{-1}h_1)}&\alpha_{\phi_2(h_2^{-1}h_2)}&\dots &\alpha_{\phi_2(h_2^{-1}h_6)}\\
\alpha_{\phi_2(h_3^{-1}h_1)}&\alpha_{\phi_2(h_3^{-1}h_2)}&\dots &\alpha_{\phi_2(h_3^{-1}h_6)}\\
\vdots & \vdots & \vdots & \vdots \\
\alpha_{\phi_2(h_6^{-1}h_1)}&\alpha_{\phi_2(h_6^{-1}h_2)}&\dots &\alpha_{\phi_2(h_6^{-1}h_6)}\\
\end{pmatrix}$$
with 
$$\phi_2 : h_j^{-1}h_k \xrightarrow{\phi_2} g_1^{-1}g_i$$
$$ \text{for when} $$
$$ j=2: \ k=1, i=7; \ k=2, i=8; \ \dots \ k=6, i=12,$$
$$ j=3: \ k=1, i=7; \ k=2, i=8; \ \dots \ k=6, i=12,$$
$$ \vdots $$
$$ j=6: \ k=1, i=7; \ k=2, i=8; \ \dots \ k=6, i=12,$$

$$A_3'=\begin{pmatrix}
\alpha_{g_1^{-1}g_{13}}&\alpha_{g_1^{-1}g_{14}}&\dots &\alpha_{g_1^{-1}g_{18}}\\
\alpha_{\phi_3(h_2^{-1}h_1)}&\alpha_{\phi_3(h_2^{-1}h_2)}&\dots &\alpha_{\phi_3(h_2^{-1}h_6)}\\
\alpha_{\phi_3(h_3^{-1}h_1)}&\alpha_{\phi_3(h_3^{-1}h_2)}&\dots &\alpha_{\phi_3(h_3^{-1}h_6)}\\
\vdots & \vdots & \vdots & \vdots \\
\alpha_{\phi_3(h_6^{-1}h_1)}&\alpha_{\phi_3(h_6^{-1}h_2)}&\dots &\alpha_{\phi_3(h_6^{-1}h_6)}\\
\end{pmatrix}$$
with 
$$\phi_3 : h_j^{-1}h_k \xrightarrow{\phi_3} g_1^{-1}g_i$$
$$ \text{for when} $$
$$ j=2: \ k=1, i=13; \ k=2, i=14; \ \dots \ k=6, i=18,$$
$$ j=3: \ k=1, i=13; \ k=2, i=14; \ \dots \ k=6, i=18,$$
$$ \vdots $$
$$ j=6: \ k=1, i=13; \ k=2, i=14; \ \dots \ k=6, i=18,$$

$$A_4'=\begin{pmatrix}
\alpha_{g_7^{-1}g_1}&\alpha_{g_7^{-1}g_2}&\dots &\alpha_{g_7^{-1}g_6}\\
\alpha_{\phi_4(h_2^{-1}h_1)}&\alpha_{\phi_4(h_2^{-1}h_2)}&\dots &\alpha_{\phi_4(h_2^{-1}h_6)}\\
\alpha_{\phi_4(h_3^{-1}h_1)}&\alpha_{\phi_4(h_3^{-1}h_2)}&\dots &\alpha_{\phi_4(h_3^{-1}h_6)}\\
\vdots & \vdots & \vdots & \vdots \\
\alpha_{\phi_4(h_6^{-1}h_1)}&\alpha_{\phi_4(h_6^{-1}h_2)}&\dots &\alpha_{\phi_4(h_6^{-1}h_6)}\\
\end{pmatrix}$$
with 
$$\phi_4 : h_j^{-1}h_i \xrightarrow{\phi_4} g_7^{-1}g_i$$
$$ \text{for when} $$
$$ j=2: \ i=1,2,\dots,6$$
$$ j=3: \ i=1,2,\dots,6$$
$$ \vdots $$
$$ j=6: \ i=1,2,\dots,6,$$

$$A_5'=\begin{pmatrix}
\alpha_{g_7^{-1}g_7}&\alpha_{g_7^{-1}g_8}&\dots &\alpha_{g_7^{-1}g_{12}}\\
\alpha_{\phi_5(h_2^{-1}h_1)}&\alpha_{\phi_5(h_2^{-1}h_2)}&\dots &\alpha_{\phi_5(h_2^{-1}h_6)}\\
\alpha_{\phi_5(h_3^{-1}h_1)}&\alpha_{\phi_5(h_3^{-1}h_2)}&\dots &\alpha_{\phi_5(h_3^{-1}h_6)}\\
\vdots & \vdots & \vdots & \vdots \\
\alpha_{\phi_5(h_6^{-1}h_1)}&\alpha_{\phi_5(h_6^{-1}h_2)}&\dots &\alpha_{\phi_5(h_6^{-1}h_6)}\\
\end{pmatrix}$$
with 
$$\phi_5 : h_j^{-1}h_k \xrightarrow{\phi_5} g_7^{-1}g_i$$
$$ \text{for when} $$
$$ j=2: \ k=1, i=7; \ k=2, i=8; \ \dots \ k=6, i=12,$$
$$ j=3: \ k=1, i=7; \ k=2, i=8; \ \dots \ k=6, i=12,$$
$$ \vdots $$
$$ j=6: \ k=1, i=7; \ k=2, i=8; \ \dots \ k=6, i=12,$$

$$A_6'=\begin{pmatrix}
\alpha_{g_7^{-1}g_{13}}&\alpha_{g_7^{-1}g_{14}}&\dots &\alpha_{g_7^{-1}g_{18}}\\
\alpha_{\phi_6(h_2^{-1}h_1)}&\alpha_{\phi_6(h_2^{-1}h_2)}&\dots &\alpha_{\phi_6(h_2^{-1}h_6)}\\
\alpha_{\phi_6(h_3^{-1}h_1)}&\alpha_{\phi_6(h_3^{-1}h_2)}&\dots &\alpha_{\phi_6(h_3^{-1}h_6)}\\
\vdots & \vdots & \vdots & \vdots \\
\alpha_{\phi_6(h_6^{-1}h_1)}&\alpha_{\phi_6(h_6^{-1}h_2)}&\dots &\alpha_{\phi_6(h_6^{-1}h_6)}\\
\end{pmatrix}$$
with 
$$\phi_6 : h_j^{-1}h_k \xrightarrow{\phi_6} g_7^{-1}g_i$$
$$ \text{for when} $$
$$ j=2: \ k=1, i=13; \ k=2, i=14; \ \dots \ k=6, i=18,$$
$$ j=3: \ k=1, i=13; \ k=2, i=14; \ \dots \ k=6, i=18,$$
$$ \vdots $$
$$ j=6: \ k=1, i=13; \ k=2, i=14; \ \dots \ k=6, i=18,$$

$$A_7'=\begin{pmatrix}
\alpha_{g_{13}^{-1}g_1}&\alpha_{g_{13}^{-1}g_2}&\dots &\alpha_{g_{13}^{-1}g_6}\\
\alpha_{\phi_7(h_2^{-1}h_1)}&\alpha_{\phi_7(h_2^{-1}h_2)}&\dots &\alpha_{\phi_7(h_2^{-1}h_6)}\\
\alpha_{\phi_7(h_3^{-1}h_1)}&\alpha_{\phi_7(h_3^{-1}h_2)}&\dots &\alpha_{\phi_7(h_3^{-1}h_6)}\\
\vdots & \vdots & \vdots & \vdots \\
\alpha_{\phi_7(h_6^{-1}h_1)}&\alpha_{\phi_7(h_6^{-1}h_2)}&\dots &\alpha_{\phi_7(h_6^{-1}h_6)}\\
\end{pmatrix}$$
with 
$$\phi_7 : h_j^{-1}h_i \xrightarrow{\phi_7} g_{13}^{-1}g_i$$
$$ \text{for when} $$
$$ j=2: \ i=1,2,\dots,6$$
$$ j=3: \ i=1,2,\dots,6$$
$$ \vdots $$
$$ j=6: \ i=1,2,\dots,6,$$

$$A_8'=\begin{pmatrix}
\alpha_{g_{13}^{-1}g_7}&\alpha_{g_{13}^{-1}g_8}&\dots &\alpha_{g_{13}^{-1}g_{12}}\\
\alpha_{\phi_8(h_2^{-1}h_1)}&\alpha_{\phi_8(h_2^{-1}h_2)}&\dots &\alpha_{\phi_8(h_2^{-1}h_6)}\\
\alpha_{\phi_8(h_3^{-1}h_1)}&\alpha_{\phi_8(h_3^{-1}h_2)}&\dots &\alpha_{\phi_8(h_3^{-1}h_6)}\\
\vdots & \vdots & \vdots & \vdots \\
\alpha_{\phi_8(h_6^{-1}h_1)}&\alpha_{\phi_8(h_6^{-1}h_2)}&\dots &\alpha_{\phi_8(h_6^{-1}h_6)}\\
\end{pmatrix}$$
with 
$$\phi_8 : h_j^{-1}h_k \xrightarrow{\phi_8} g_{13}^{-1}g_i$$
$$ \text{for when} $$
$$ j=2: \ k=1, i=7; \ k=2, i=8; \ \dots \ k=6, i=12,$$
$$ j=3: \ k=1, i=7; \ k=2, i=8; \ \dots \ k=6, i=12,$$
$$ \vdots $$
$$ j=6: \ k=1, i=7; \ k=2, i=8; \ \dots \ k=6, i=12,$$
and
$$A_9'=\begin{pmatrix}
\alpha_{g_{13}^{-1}g_{13}}&\alpha_{g_{13}^{-1}g_{14}}&\dots &\alpha_{g_{13}^{-1}g_{18}}\\
\alpha_{\phi_9(h_2^{-1}h_1)}&\alpha_{\phi_9(h_2^{-1}h_2)}&\dots &\alpha_{\phi_9(h_2^{-1}h_6)}\\
\alpha_{\phi_9(h_3^{-1}h_1)}&\alpha_{\phi_9(h_3^{-1}h_2)}&\dots &\alpha_{\phi_9(h_3^{-1}h_6)}\\
\vdots & \vdots & \vdots & \vdots \\
\alpha_{\phi_9(h_6^{-1}h_1)}&\alpha_{\phi_9(h_6^{-1}h_2)}&\dots &\alpha_{\phi_9(h_6^{-1}h_6)}\\
\end{pmatrix}$$
with 
$$\phi_9 : h_j^{-1}h_k \xrightarrow{\phi_9} g_{13}^{-1}g_i$$
$$ \text{for when} $$
$$ j=2: \ k=1, i=13; \ k=2, i=14; \ \dots \ k=6, i=18,$$
$$ j=3: \ k=1, i=13; \ k=2, i=14; \ \dots \ k=6, i=18,$$
$$ \vdots $$
$$ j=6: \ k=1, i=13; \ k=2, i=14; \ \dots \ k=6, i=18.$$
This results in a composite matrix over $R$ of the following form:
$$\Omega(v_2)=\begin{pmatrix}
A_1'&A_2'&A_3'\\
A_4'&A_5'&A_6'\\
A_7'&A_8'&A_9'
\end{pmatrix}=\begin{pmatrix}
B&C&D\\
D&B&C\\
C&D&B
\end{pmatrix}=$$
$$=\begin{pmatrix}
B_1&B_2&C_1&C_2&D_1&D_2\\
B_2^T&B_1^T&C_2^T&C_1^T&D_2^T&D_1^T\\
D_1&D_2&B_1&B_2&C_1&C_2\\
D_2^T&D_1^T&B_2^T&B_1^T&C_2^T&C_1^T\\
C_1&C_2&D_1&D_2&B_1&B_2\\
C_2^T&C_1^T&D_2^T&D_1^T&B_2^T&B_1^T
\end{pmatrix},$$
where
$$B_1=circ(\alpha_1,\alpha_2,\alpha_3), B_2=circ(\alpha_4,\alpha_5,\alpha_6), C_1=circ(\alpha_7,\alpha_{8},\alpha_{9}),$$
$$C_2=circ(\alpha_{10},\alpha_{11},\alpha_{12}), D_1=circ(\alpha_{13},\alpha_{14},\alpha_{15}), D_2=circ(\alpha_{16},\alpha_{17},\alpha_{18}).$$

\item[3.] Let $G=\langle x,y \ | \ x^6=y^3=1, xy=yx \rangle \cong C_3 \times C_6.$ Let $v_3=\sum_{i=0}^5 \sum_{j=0}^2 \alpha_{1+i+6j}x^iy^j \in R(C_3 \times C_6).$  Also, let $H=\langle a \ | \ a^{6}=1 \rangle \cong C_{3,2}.$ We now define the composite matrix over $R$ as
$$\Omega(v_3)=\begin{pmatrix}
A_1'&A_2'&A_3'\\
A_4'&A_5'&A_6'\\
A_7'&A_8'&A_9'
\end{pmatrix}$$
where
$$A_1'=\begin{pmatrix}
\alpha_{g_1^{-1}g_1}&\alpha_{g_1^{-1}g_2}&\dots &\alpha_{g_1^{-1}g_6}\\
\alpha_{\phi_1(h_2^{-1}h_1)}&\alpha_{\phi_1(h_2^{-1}h_2)}&\dots &\alpha_{\phi_1(h_2^{-1}h_6)}\\
\alpha_{\phi_1(h_3^{-1}h_1)}&\alpha_{\phi_1(h_3^{-1}h_2)}&\dots &\alpha_{\phi_1(h_3^{-1}h_6)}\\
\vdots & \vdots & \vdots & \vdots \\
\alpha_{\phi_1(h_6^{-1}h_1)}&\alpha_{\phi_1(h_6^{-1}h_2)}&\dots &\alpha_{\phi_1(h_6^{-1}h_6)}\\
\end{pmatrix}$$
with 
$$\phi_1 : h_j^{-1}h_i \xrightarrow{\phi_1} g_1^{-1}g_i$$
$$ \text{for when} $$
$$ j=2: \ i=1,2,\dots,6$$
$$ j=3: \ i=1,2,\dots,6$$
$$ \vdots $$
$$ j=6: \ i=1,2,\dots,6,$$

$$A_2'=\begin{pmatrix}
\alpha_{g_1^{-1}g_7}&\alpha_{g_1^{-1}g_8}&\dots &\alpha_{g_1^{-1}g_{12}}\\
\alpha_{\phi_2(h_2^{-1}h_1)}&\alpha_{\phi_2(h_2^{-1}h_2)}&\dots &\alpha_{\phi_2(h_2^{-1}h_6)}\\
\alpha_{\phi_2(h_3^{-1}h_1)}&\alpha_{\phi_2(h_3^{-1}h_2)}&\dots &\alpha_{\phi_2(h_3^{-1}h_6)}\\
\vdots & \vdots & \vdots & \vdots \\
\alpha_{\phi_2(h_6^{-1}h_1)}&\alpha_{\phi_2(h_6^{-1}h_2)}&\dots &\alpha_{\phi_2(h_6^{-1}h_6)}\\
\end{pmatrix}$$
with 
$$\phi_2 : h_j^{-1}h_k \xrightarrow{\phi_2} g_1^{-1}g_i$$
$$ \text{for when} $$
$$ j=2: \ k=1, i=7; \ k=2, i=8; \ \dots \ k=6, i=12,$$
$$ j=3: \ k=1, i=7; \ k=2, i=8; \ \dots \ k=6, i=12,$$
$$ \vdots $$
$$ j=6: \ k=1, i=7; \ k=2, i=8; \ \dots \ k=6, i=12,$$

$$A_3'=\begin{pmatrix}
\alpha_{g_1^{-1}g_{13}}&\alpha_{g_1^{-1}g_{14}}&\dots &\alpha_{g_1^{-1}g_{18}}\\
\alpha_{\phi_3(h_2^{-1}h_1)}&\alpha_{\phi_3(h_2^{-1}h_2)}&\dots &\alpha_{\phi_3(h_2^{-1}h_6)}\\
\alpha_{\phi_3(h_3^{-1}h_1)}&\alpha_{\phi_3(h_3^{-1}h_2)}&\dots &\alpha_{\phi_3(h_3^{-1}h_6)}\\
\vdots & \vdots & \vdots & \vdots \\
\alpha_{\phi_3(h_6^{-1}h_1)}&\alpha_{\phi_3(h_6^{-1}h_2)}&\dots &\alpha_{\phi_3(h_6^{-1}h_6)}\\
\end{pmatrix}$$
with 
$$\phi_3 : h_j^{-1}h_k \xrightarrow{\phi_3} g_1^{-1}g_i$$
$$ \text{for when} $$
$$ j=2: \ k=1, i=13; \ k=2, i=14; \ \dots \ k=6, i=18,$$
$$ j=3: \ k=1, i=13; \ k=2, i=14; \ \dots \ k=6, i=18,$$
$$ \vdots $$
$$ j=6: \ k=1, i=13; \ k=2, i=14; \ \dots \ k=6, i=18,$$

$$A_4'=\begin{pmatrix}
\alpha_{g_7^{-1}g_1}&\alpha_{g_7^{-1}g_2}&\dots &\alpha_{g_7^{-1}g_6}\\
\alpha_{\phi_4(h_2^{-1}h_1)}&\alpha_{\phi_4(h_2^{-1}h_2)}&\dots &\alpha_{\phi_4(h_2^{-1}h_6)}\\
\alpha_{\phi_4(h_3^{-1}h_1)}&\alpha_{\phi_4(h_3^{-1}h_2)}&\dots &\alpha_{\phi_4(h_3^{-1}h_6)}\\
\vdots & \vdots & \vdots & \vdots \\
\alpha_{\phi_4(h_6^{-1}h_1)}&\alpha_{\phi_4(h_6^{-1}h_2)}&\dots &\alpha_{\phi_4(h_6^{-1}h_6)}\\
\end{pmatrix}$$
with 
$$\phi_4 : h_j^{-1}h_i \xrightarrow{\phi_4} g_7^{-1}g_i$$
$$ \text{for when} $$
$$ j=2: \ i=1,2,\dots,6$$
$$ j=3: \ i=1,2,\dots,6$$
$$ \vdots $$
$$ j=6: \ i=1,2,\dots,6,$$

$$A_5'=\begin{pmatrix}
\alpha_{g_7^{-1}g_7}&\alpha_{g_7^{-1}g_8}&\dots &\alpha_{g_7^{-1}g_{12}}\\
\alpha_{\phi_5(h_2^{-1}h_1)}&\alpha_{\phi_5(h_2^{-1}h_2)}&\dots &\alpha_{\phi_5(h_2^{-1}h_6)}\\
\alpha_{\phi_5(h_3^{-1}h_1)}&\alpha_{\phi_5(h_3^{-1}h_2)}&\dots &\alpha_{\phi_5(h_3^{-1}h_6)}\\
\vdots & \vdots & \vdots & \vdots \\
\alpha_{\phi_5(h_6^{-1}h_1)}&\alpha_{\phi_5(h_6^{-1}h_2)}&\dots &\alpha_{\phi_5(h_6^{-1}h_6)}\\
\end{pmatrix}$$
with 
$$\phi_5 : h_j^{-1}h_k \xrightarrow{\phi_5} g_7^{-1}g_i$$
$$ \text{for when} $$
$$ j=2: \ k=1, i=7; \ k=2, i=8; \ \dots \ k=6, i=12,$$
$$ j=3: \ k=1, i=7; \ k=2, i=8; \ \dots \ k=6, i=12,$$
$$ \vdots $$
$$ j=6: \ k=1, i=7; \ k=2, i=8; \ \dots \ k=6, i=12,$$

$$A_6'=\begin{pmatrix}
\alpha_{g_7^{-1}g_{13}}&\alpha_{g_7^{-1}g_{14}}&\dots &\alpha_{g_7^{-1}g_{18}}\\
\alpha_{\phi_6(h_2^{-1}h_1)}&\alpha_{\phi_6(h_2^{-1}h_2)}&\dots &\alpha_{\phi_6(h_2^{-1}h_6)}\\
\alpha_{\phi_6(h_3^{-1}h_1)}&\alpha_{\phi_6(h_3^{-1}h_2)}&\dots &\alpha_{\phi_6(h_3^{-1}h_6)}\\
\vdots & \vdots & \vdots & \vdots \\
\alpha_{\phi_6(h_6^{-1}h_1)}&\alpha_{\phi_6(h_6^{-1}h_2)}&\dots &\alpha_{\phi_6(h_6^{-1}h_6)}\\
\end{pmatrix}$$
with 
$$\phi_6 : h_j^{-1}h_k \xrightarrow{\phi_6} g_7^{-1}g_i$$
$$ \text{for when} $$
$$ j=2: \ k=1, i=13; \ k=2, i=14; \ \dots \ k=6, i=18,$$
$$ j=3: \ k=1, i=13; \ k=2, i=14; \ \dots \ k=6, i=18,$$
$$ \vdots $$
$$ j=6: \ k=1, i=13; \ k=2, i=14; \ \dots \ k=6, i=18,$$

$$A_7'=\begin{pmatrix}
\alpha_{g_{13}^{-1}g_1}&\alpha_{g_{13}^{-1}g_2}&\dots &\alpha_{g_{13}^{-1}g_6}\\
\alpha_{\phi_7(h_2^{-1}h_1)}&\alpha_{\phi_7(h_2^{-1}h_2)}&\dots &\alpha_{\phi_7(h_2^{-1}h_6)}\\
\alpha_{\phi_7(h_3^{-1}h_1)}&\alpha_{\phi_7(h_3^{-1}h_2)}&\dots &\alpha_{\phi_7(h_3^{-1}h_6)}\\
\vdots & \vdots & \vdots & \vdots \\
\alpha_{\phi_7(h_6^{-1}h_1)}&\alpha_{\phi_7(h_6^{-1}h_2)}&\dots &\alpha_{\phi_7(h_6^{-1}h_6)}\\
\end{pmatrix}$$
with 
$$\phi_7 : h_j^{-1}h_i \xrightarrow{\phi_7} g_{13}^{-1}g_i$$
$$ \text{for when} $$
$$ j=2: \ i=1,2,\dots,6$$
$$ j=3: \ i=1,2,\dots,6$$
$$ \vdots $$
$$ j=6: \ i=1,2,\dots,6,$$

$$A_8'=\begin{pmatrix}
\alpha_{g_{13}^{-1}g_7}&\alpha_{g_{13}^{-1}g_8}&\dots &\alpha_{g_{13}^{-1}g_{12}}\\
\alpha_{\phi_8(h_2^{-1}h_1)}&\alpha_{\phi_8(h_2^{-1}h_2)}&\dots &\alpha_{\phi_8(h_2^{-1}h_6)}\\
\alpha_{\phi_8(h_3^{-1}h_1)}&\alpha_{\phi_8(h_3^{-1}h_2)}&\dots &\alpha_{\phi_8(h_3^{-1}h_6)}\\
\vdots & \vdots & \vdots & \vdots \\
\alpha_{\phi_8(h_6^{-1}h_1)}&\alpha_{\phi_8(h_6^{-1}h_2)}&\dots &\alpha_{\phi_8(h_6^{-1}h_6)}\\
\end{pmatrix}$$
with 
$$\phi_8 : h_j^{-1}h_k \xrightarrow{\phi_8} g_{13}^{-1}g_i$$
$$ \text{for when} $$
$$ j=2: \ k=1, i=7; \ k=2, i=8; \ \dots \ k=6, i=12,$$
$$ j=3: \ k=1, i=7; \ k=2, i=8; \ \dots \ k=6, i=12,$$
$$ \vdots $$
$$ j=6: \ k=1, i=7; \ k=2, i=8; \ \dots \ k=6, i=12,$$
and
$$A_9'=\begin{pmatrix}
\alpha_{g_{13}^{-1}g_{13}}&\alpha_{g_{13}^{-1}g_{14}}&\dots &\alpha_{g_{13}^{-1}g_{18}}\\
\alpha_{\phi_9(h_2^{-1}h_1)}&\alpha_{\phi_9(h_2^{-1}h_2)}&\dots &\alpha_{\phi_9(h_2^{-1}h_6)}\\
\alpha_{\phi_9(h_3^{-1}h_1)}&\alpha_{\phi_9(h_3^{-1}h_2)}&\dots &\alpha_{\phi_9(h_3^{-1}h_6)}\\
\vdots & \vdots & \vdots & \vdots \\
\alpha_{\phi_9(h_6^{-1}h_1)}&\alpha_{\phi_9(h_6^{-1}h_2)}&\dots &\alpha_{\phi_9(h_6^{-1}h_6)}\\
\end{pmatrix}$$
with 
$$\phi_9 : h_j^{-1}h_k \xrightarrow{\phi_9} g_{13}^{-1}g_i$$
$$ \text{for when} $$
$$ j=2: \ k=1, i=13; \ k=2, i=14; \ \dots \ k=6, i=18,$$
$$ j=3: \ k=1, i=13; \ k=2, i=14; \ \dots \ k=6, i=18,$$
$$ \vdots $$
$$ j=6: \ k=1, i=13; \ k=2, i=14; \ \dots \ k=6, i=18.$$
This results in a composite matrix over $R$ of the following form:
$$\Omega(v_3)=\begin{pmatrix}
A_1'&A_2'&A_3'\\
A_4'&A_5'&A_6'\\
A_7'&A_8'&A_9'
\end{pmatrix}=\begin{pmatrix}
B&C&D\\
D&B&C\\
C&D&B
\end{pmatrix}=$$
$$=\begin{pmatrix}
B_1&B_2&C_1&C_2&D_1&D_2\\
B_2'&B_1&C_2'&C_1&D_2'&D_1\\
D_1&D_2&B_1&B_2&C_1&C_2\\
D_2'&D_1&B_2'&B_1&C_2'&C_1\\
C_1&C_2&D_1&D_2&B_1&B_2\\
C_2'&C_1&D_2'&D_1&B_2'&B_1
\end{pmatrix},$$
where
$$B_1=circ(\alpha_1,\alpha_2,\alpha_3), B_2=circ(\alpha_4,\alpha_5,\alpha_6), B_2'=circ(\alpha_6,\alpha_4,\alpha_5),$$
$$C_1=circ(\alpha_7,\alpha_8,\alpha_9), C_2=circ(\alpha_{10},\alpha_{11},\alpha_{12}), C_2'=circ(\alpha_{12},\alpha_{10},\alpha_{11}),$$
$$D_1=circ(\alpha_{13},\alpha_{14},\alpha_{15}), D_2=circ(\alpha_{16},\alpha_{17},\alpha_{18}), D_2'=circ(\alpha_{18},\alpha_{16},\alpha_{17}).$$

\end{enumerate}

\section{Generator Matrices}\label{genmatrices}

In this section, we consider generator matrices of the form $[I \ | \ \Omega(v_i)],$ where $I$ is the identity matrix and $\Omega(v_i)$ with $i=\{1,2,3\}$ are the composite matrices from the previous section. For each, we show under what conditions such generator matrix produces self-dual codes over the ring $R.$ We assume that the ring $R$ has characteristic 2.

\begin{theorem}\label{thm1}
The generator matrix
\begin{equation}
\begin{bmatrix}
I&|&\Omega(v_1)
\end{bmatrix}
\end{equation}
where $\Omega(v_1)$ is the composite matrix defined in previous section, generates a self-dual code over the ring $R$ if and only if the following hold in $R:$
\begin{equation}
BB^T+CC^T=I_9,
\end{equation}
\begin{equation}
BD^T+CE^T=\mathbf{0},
\end{equation}
\begin{equation}
DB^T+EC^T=\mathbf{0},
\end{equation}
\begin{equation}
DD^T+EE^T=I_9.
\end{equation}
\begin{proof}
Follows from the standard proof that $(I_m \ | \ A)$ generates a self-dual of length $2m$ if and only if $AA^T=I_m.$
\end{proof}
\end{theorem}

\begin{theorem}\label{thm2}
The generator matrix
\begin{equation}
\begin{bmatrix}
I&|&\Omega(v_2)
\end{bmatrix}
\end{equation}
where $\Omega(v_2)$ is the composite matrix defined in previous section, generates a self-dual code over the ring $R$ if and only if the following hold in $R:$
\begin{equation}
BB^T+CC^T+DD^T=I_6,
\end{equation}
\begin{equation}
BD^T+CB^T+DC^T=\mathbf{0},
\end{equation}
\begin{equation}
BC^T+CD^T+DB^T=\mathbf{0}.
\end{equation}
\begin{proof}
Follows from the standard proof that $(I_m \ | \ A)$ generates a self-dual of length $2m$ if and only if $AA^T=I_m.$
\end{proof}
\end{theorem}

\begin{theorem}\label{thm3}
The generator matrix
\begin{equation}
\begin{bmatrix}
I&|&\Omega(v_3)
\end{bmatrix}
\end{equation}
where $\Omega(v_3)$ is the composite matrix defined in previous section, generates a self-dual code over the ring $R$ if and only if the following hold in $R:$
\begin{equation}
BB^T+CC^T+DD^T=I_6,
\end{equation}
\begin{equation}
BD^T+CB^T+DC^T=\mathbf{0},
\end{equation}
\begin{equation}
BC^T+CD^T+DB^T=\mathbf{0}.
\end{equation}
\begin{proof}
Follows from the standard proof that $(I_m \ | \ A)$ generates a self-dual of length $2m$ if and only if $AA^T=I_m.$
\end{proof}
\end{theorem}

\section{New Type I Binary Self-Dual Codes of length 72}

In this section, we search for binary self-dual codes by employing the generator matrices defined in Section~\ref{genmatrices}. In particular, we search for binary self-dual codes with parameters $[36, 18, 6 \ \text{or} \ 8]$ which we then lift to the ring $\mathbb{F}_2+u\mathbb{F}_2$ to obtain self-dual codes of length $36$ whose binary images are self-dual codes with parameters $[72,36,12].$

The possible weight enumerators for Type~I $[72,36,12]$ codes are as follows (\cite{DoughertyIV}):
$$W_{72,1}=1+2\beta y^{12}+(8640-64\gamma)y^{14}+(124281-24\beta+384\gamma)y^{16}+\dots$$
$$W_{72,2}=1+2 \beta y^{12}+(7616-64 \gamma)y^{14}+(134521-24 \beta+384 \gamma)y^{16}+\dots$$
where $\beta$ and $\gamma$ are parameters.

Many codes for different values of $\beta$ and $\gamma$ have been constructed in \cite{Bouyukliev1, Dontcheva1, DoughertyIV, Dougherty2, Gulliver1, Yankov2, Kaya1, Korban1, Yildiz1, Yankov1, Zhdanov1, Zhdanov2}. For an up-to-date list of all known Type~I and Type~II binary self-dual codes with parameters $[72,36,12]$ please see \cite{selfdual72}.

All the upcoming computational results were obtained by performing searches in the software package MAGMA (\cite{MAGMA}).

\begin{enumerate}
\item[1.] Here, we employ the generator matrix $[I \ | \ \Omega(v_1)]$ to search for binary self-dual codes of length 36. Since the matrix $\Omega(v_1)$ is fully defined by the first row, we only list the first rows of the matrices $B$ and $C$ which we label as $r_B$ and $r_C$ respectively. We summarise the results in the table below.

\begin{table}[h!]
\caption{Type I $[36,18,6-8]$ Codes from Theorem~\ref{thm1}}
\centering
\resizebox{1\textwidth}{!}{\begin{minipage}{\textwidth}
\centering
\begin{tabular}{ccccc}
\hline
      & Type        & $r_B$                 & $r_C$                 & $|Aut(C_i)|$ \\ \hline
$C_1$ & $[36,18,6]$ & $(0,0,0,0,0,1,0,1,1)$ & $(1,0,1,1,1,0,1,0,1)$ & $2^2 \cdot 3^2$         \\ \hline
$C_2$ & $[36,18,6]$ & $(0,0,0,0,1,1,0,1,1)$ & $(1,0,0,1,0,1,1,1,0)$ & $2^2 \cdot 3^2$         \\ \hline
$C_3$ & $[36,18,8]$ & $(0,0,1,0,0,1,0,0,1)$ & $(1,0,0,1,1,0,1,1,1)$ & $2^2 \cdot 3^2$         \\ \hline
$C_4$ & $[36,18,8]$  & $(0,1,0,0,0,1,0,1,1)$ & $(1,0,0,1,0,0,1,1,1)$ & $2^2 \cdot 3^2$         \\ \hline
\end{tabular}
\end{minipage}}
\end{table}

We now apply the $R_1$-lift to each code from the above table to obtain codes whose binary images are self-dual codes with parameters $[72,36,12].$ We only list codes that have not been known in the literature before.

\begin{table}[h!]
\caption{New Type I $[72,36,12]$ Codes from $R_1$-lift of $C_1$}
\resizebox{0.8\textwidth}{!}{\begin{minipage}{\textwidth}
\centering
\begin{tabular}{ccccccc}
\hline
      & Type       & $r_B$                                   & $r_C$                                   & $\gamma$ & $\beta$ & $|Aut(\mathcal{C}_i)|$ \\ \hline
$\mathcal{C}_1$ & $W_{72,1}$ & $(u,0,u,u,u,1,u,u + 1,1)$ & $(1,0,1,1,u + 1,0,1,u,1)$ & $0$      & $192$   & $36$         \\
\hline
$\mathcal{C}_2$ & $W_{72,1}$ & $(u,0,0,u,0,1,u,1,1)$ & $(1,0,u + 1,1,u + 1,0,1,u,u + 1)$ & $0$      & $198$   & $36$         \\ \hline
$\mathcal{C}_3$ & $W_{72,1}$ & $(u,u,0,u,u,1,u,1,u + 1)$ & $(1,u,u + 1,1,u + 1,0,1,0,u + 1)$ & $0$      & $336$   & $36$         \\ \hline
$\mathcal{C}_4$ & $W_{72,1}$ & $(0,u,0,0,0,1,0,1,u + 1)$ & $(1,u,u + 1,1,u + 1,0,1,0,u + 1)$ & $18$      & $234$   & $36$         \\ \hline
$\mathcal{C}_5$ & $W_{72,1}$ & $(u,u,0,u,0,1,u,u + 1,1)$ & $(1,u,u + 1,1,1,u,1,0,u + 1)$ & $18$      & $345$   & $36$         \\ \hline
$\mathcal{C}_6$ & $W_{72,1}$ & $(0,u,0,0,0,1,0,u + 1,1)$ & $(1,u,1,1,u + 1,u,1,u,1)$ & $18$      & $378$   & $36$         \\ \hline
$\mathcal{C}_7$ & $W_{72,1}$ & $(u,u,u,u,0,1,u,u + 1,u + 1)$ & $(1,u,1,1,u + 1,0,1,0,1)$ & $18$      & $396$   & $36$         \\ \hline
$\mathcal{C}_8$ & $W_{72,1}$ & $(0,0,u,0,u,1,0,1,u + 1)$ & $(1,u,1,1,1,0,1,u,1)$ & $18$      & $441$   & $36$         \\ \hline
$\mathcal{C}_9$ & $W_{72,1}$ & $(u,u,0,u,0,1,u,1,u + 1)$ & $(1,u,1,1,1,0,1,u,1)$ & $18$      & $453$   & $36$         \\ \hline
\end{tabular}
\end{minipage}}
\end{table}

\begin{table}[h!]
\caption{New Type I $[72,36,12]$ Codes from $R_1$-lift of $C_2$}
\resizebox{0.8\textwidth}{!}{\begin{minipage}{\textwidth}
\centering
\begin{tabular}{ccccccc}
\hline
      & Type       & $r_B$                                   & $r_C$                                   & $\gamma$ & $\beta$ & $|Aut(\mathcal{C}_i)|$ \\ \hline
$\mathcal{C}_{10}$ & $W_{72,1}$ & $(0,u,0,0,1,1,0,1,u + 1)$ & $(1,u,u,1,u,u + 1,1,1,0)$ & $0$      & $219$   & $36$         \\
\hline
$\mathcal{C}_{11}$ & $W_{72,1}$ & $(u,0,u,u,1,1,u,u + 1,u + 1)$ & $(1,u,0,1,0,1,1,u + 1,u)$ & $0$      & $345$   & $36$         \\ \hline
$\mathcal{C}_{12}$ & $W_{72,1}$ & $(0,0,u,0,1,1,0,1,u + 1)$ & $(1,u,0,1,0,1,1,1,0)$ & $0$      & $408$   & $36$         \\ \hline
$\mathcal{C}_{13}$ & $W_{72,1}$ & $(u,0,0,u,1,u + 1,u,u + 1,u + 1)$ & $(1,0,u,1,0,u + 1,1,u + 1,0)$ & $18$      & $261$   & $36$         \\ \hline
$\mathcal{C}_{14}$ & $W_{72,1}$ & $(u,u,u,u,1,1,u,1,u + 1)$ & $(1,0,0,1,u,1,1,u + 1,0)$ & $18$      & $270$   & $36$         \\ \hline
$\mathcal{C}_{15}$ & $W_{72,1}$ & $(u,0,u,u,1,u + 1,u,1,u + 1)$ & $(1,0,u,1,u,1,1,u + 1,0)$ & $18$      & $357$   & $36$         \\ \hline
\end{tabular}
\end{minipage}}
\end{table}

\begin{table}[h!]
\caption{New Type I $[72,36,12]$ Codes from $R_1$-lift of $C_3$}
\resizebox{0.8\textwidth}{!}{\begin{minipage}{\textwidth}
\centering
\begin{tabular}{ccccccc}
\hline
      & Type       & $r_B$                                   & $r_C$                                   & $\gamma$ & $\beta$ & $|Aut(\mathcal{C}_i)|$ \\ \hline
$\mathcal{C}_{16}$ & $W_{72,1}$ & $(u,u,1,u,0,1,u,0,1)$ & $(1,u,0,1,u + 1,u,1,1,u + 1)$ & $0$      & $120$   & $36$         \\
\hline
$\mathcal{C}_{17}$ & $W_{72,1}$ & $(u,u,1,u,0,1,u,u,1)$ & $(1,0,u,1,1,0,1,1,u + 1)$ & $0$      & $282$   & $36$         \\ \hline
$\mathcal{C}_{18}$ & $W_{72,1}$ & $(u,u,1,u,0,u + 1,u,0,1)$ & $(1,u,0,1,u + 1,u,1,1,u + 1)$ & $0$      & $300$   & $36$         \\ \hline
$\mathcal{C}_{19}$ & $W_{72,1}$ & $(u,u,1,u,0,u + 1,u,0,1)$ & $(1,u,u,1,1,0,1,1,1)$ & $18$      & $336$   & $36$         \\ \hline
$\mathcal{C}_{20}$ & $W_{72,1}$ & $(u,0,1,u,0,1,u,u,1)$ & $(1,0,0,1,1,0,1,u + 1,u + 1)$ & $36$      & $435$   & $36$         \\ \hline
\end{tabular}
\end{minipage}}
\end{table}

\begin{table}[h!]
\caption{New Type I $[72,36,12]$ Codes from $R_1$-lift of $C_4$}
\resizebox{0.8\textwidth}{!}{\begin{minipage}{\textwidth}
\centering
\begin{tabular}{ccccccc}
\hline
      & Type       & $r_B$                                   & $r_C$                                   & $\gamma$ & $\beta$ & $|Aut(\mathcal{C}_i)|$ \\ \hline
$\mathcal{C}_{21}$ & $W_{72,1}$ & $(0,1,u,0,u,1,0,u + 1,u + 1)$ & $(1,u,u,1,u,u,1,u + 1,u + 1)$ & $0$      & $366$   & $36$         \\
\hline
$\mathcal{C}_{22}$ & $W_{72,1}$ & $(u,1,u,u,u,1,u,1,u + 1)$ & $(1,u,0,1,0,0,1,u + 1,1)$ & $0$      & $372$   & $36$         \\ \hline
$\mathcal{C}_{23}$ & $W_{72,1}$ & $(0,1,u,0,u,1,0,u + 1,u + 1)$ & $(1,0,0,1,0,0,1,u + 1,u + 1)$ & $0$      & $384$   & $36$         \\ \hline
$\mathcal{C}_{24}$ & $W_{72,1}$ & $(u,1,u,u,u,1,u,1,u + 1)$ & $(1,u,u,1,u,0,1,u + 1,1)$ & $0$      & $390$   & $36$         \\ \hline
$\mathcal{C}_{25}$ & $W_{72,1}$ & $(u,1,0,u,0,1,u,u + 1,1)$ & $(1,u,0,1,0,0,1,1,u + 1)$ & $0$      & $399$   & $36$         \\ \hline
$\mathcal{C}_{26}$ & $W_{72,1}$ & $(u,1,u,u,u,1,u,u + 1,1)$ & $(1,0,u,1,0,0,1,u + 1,u + 1)$ & $18$      & $264$   & $36$         \\ \hline
$\mathcal{C}_{27}$ & $W_{72,1}$ & $(u,1,0,u,u,u + 1,u,u + 1,1)$ & $(1,0,u,1,0,u,1,1,u + 1)$ & $18$      & $285$   & $36$         \\ \hline
$\mathcal{C}_{28}$ & $W_{72,1}$ & $(0,1,u,0,0,u + 1,0,1,1)$ & $(1,u,u,1,u,0,1,1,1)$ & $18$      & $300$   & $36$         \\ \hline
\end{tabular}
\end{minipage}}
\end{table}
\newpage

\item[2.] Here, we employ the generator matrix $[I \ | \ \Omega(v_2)]$ to search for binary self-dual codes of length 36. Since the matrix $\Omega(v_2)$ is fully defined by the first row, we only list the first rows of the matrices $B, C$ and $D$ which we label as $r_B, r_C$ and $r_D$ respectively. We summarise the results in the table below

\begin{table}[h!]
\caption{Type I $[36,18,6-8]$ Codes from Theorem~\ref{thm2}}
\centering
\resizebox{1\textwidth}{!}{\begin{minipage}{\textwidth}
\centering
\begin{tabular}{cccccc}
\hline
      & Type        & $r_B$                 & $r_C$   & $r_D$              & $|Aut(C_i)|$ \\ \hline
$C_5$ & $[36,18,6]$ & $(0,0,0,0,0,1)$ & $(1,1,1,0,0,1)$ & $(1,1,1,0,1,0)$ & $2^5 \cdot 3^4 \cdot 5$         \\ \hline
$C_6$ & $[36,18,6]$ & $(0,0,0,0,1,1)$ & $(0,0,0,0,1,1)$ & $(1,1,1,1,0,1)$ & $2^5 \cdot 3^4 \cdot 5$         \\ \hline
$C_7$ & $[36,18,6]$ & $(0,0,0,0,1,1)$ & $(0,1,1,0,1,1)$ & $(0,1,1,1,0,0)$ & $2^5 \cdot 3^2$         \\ \hline
$C_8$ & $[36,18,6]$ & $(0,0,1,0,0,1)$ & $(0,0,1,1,1,0)$ & $(1,1,1,0,0,1)$ & $2^5 \cdot 3^2$         \\ \hline
\end{tabular}
\end{minipage}}
\end{table}

We now apply the $R_1$-lift to each code from the above table to obtain codes whose binary images are self-dual codes with parameters $[72,36,12].$ We only list codes that have not been known in the literature before.

\begin{table}[h!]
\caption{New Type I $[72,36,12]$ Codes from $R_1$-lift of $C_7$}
\resizebox{0.8\textwidth}{!}{\begin{minipage}{\textwidth}
\centering
\begin{tabular}{cccccccc}
\hline
      & Type       & $r_B$                                   & $r_C$                                  & $r_D$ & $\gamma$ & $\beta$ & $|Aut(\mathcal{C}_i)|$ \\ \hline
$\mathcal{C}_{29}$ & $W_{72,1}$ & $(0,0,0,u,1,1)$ & $(u,1,u + 1,u,1,1)$ & $(u,u + 1,1,u + 1,0,0)$ & $0$      & $471$   & $144$         \\
\hline
\end{tabular}
\end{minipage}}
\end{table}

\item[3.] Here, we employ the generator matrix $[I \ | \ \Omega(v_3)]$ to search for binary self-dual codes of length 36. Since the matrix $\Omega(v_3)$ is fully defined by the first row, we only list the first rows of the matrices $B, C$ and $D$ which we label as $r_B, r_C$ and $r_D$ respectively. We summarise the results in the table below

\begin{table}[h!]
\caption{Type I $[36,18,6-8]$ Codes from Theorem~\ref{thm2}}
\centering
\resizebox{1\textwidth}{!}{\begin{minipage}{\textwidth}
\centering
\begin{tabular}{cccccc}
\hline
      & Type        & $r_B$                 & $r_C$   & $r_D$              & $|Aut(C_i)|$ \\ \hline
$C_9$ & $[36,18,6]$ & $(0,0,0,0,0,1)$ & $(0,1,1,0,1,1)$ & $(1,0,1,1,0,1)$ & $2^5 \cdot 3^2$         \\ \hline
$C_{10}$ & $[36,18,6]$ & $(0,0,0,0,0,1)$ & $(1,1,1,0,0,1)$ & $(1,1,1,0,1,0)$ & $2^5 \cdot 3^4 \cdot 5$         \\ \hline
$C_{11}$ & $[36,18,6]$ & $(0,0,0,0,1,1)$ & $(0,0,0,0,1,1)$ & $(1,1,1,1,0,1)$ & $2^5 \cdot 3^4 \cdot 5$         \\ \hline
$C_{12}$ & $[36,18,6]$ & $(0,0,0,0,1,1)$ & $(0,1,1,0,0,1)$ & $(1,1,0,1,0,1)$ & $2^5 \cdot 3^2$         \\ \hline
\end{tabular}
\end{minipage}}
\end{table}

We now apply the $R_1$-lift to each code from the above table to obtain codes whose binary images are self-dual codes with parameters $[72,36,12].$ We only list codes that have not been known in the literature before.

\begin{table}[h!]
\caption{New Type I $[72,36,12]$ Codes from $R_1$-lift of $C_9$}
\resizebox{0.8\textwidth}{!}{\begin{minipage}{\textwidth}
\centering
\begin{tabular}{cccccccc}
\hline
      & Type       & $r_B$                                   & $r_C$                                  & $r_D$ & $\gamma$ & $\beta$ & $|Aut(\mathcal{C}_i)|$ \\ \hline
$\mathcal{C}_{30}$ & $W_{72,1}$ & $(0,u,u,u,u,1)$ & $(u,1,1,u,u + 1,1)$ & $(u + 1,u,u + 1,1,u,u + 1)$ & $0$      & $621$   & $432$         \\
\hline
\end{tabular}
\end{minipage}}
\end{table}

\end{enumerate}

\section{Conclusion}

In this paper, we presented three composite matrices of order $18$ derived from group rings. We formed three generator matrices which consist of the composite matrices and searched for binary self-dual codes with parameters $[36,18, 6 \ \text{or} \ 8].$ We then lifted the codes over the ring $R_1=\mathbb{F}_2+u\mathbb{F}_2$ to obtain codes whose binary images are self-dual codes with parameters $[72,36,12].$ We were able to construct $30$ Type I binary $[72,36,12]$ self-dual codes with new weight enumerators in $W_{72,1}$:
\begin{equation*}
\begin{array}{l}
(\gamma =0,\ \  \beta =\{120, 192, 198, 219, 282, 300, 336, 345, 366, 372, 384, 390, 399, 408,\\
 \quad \quad \quad \quad \quad \quad 471, 621\}), \\
(\gamma =18,\  \beta =\{234, 261, 264, 270, 285, 300, 336, 345,, 357, 378, 396, 441, 453 \}), \\
(\gamma =36,\ \beta =\{435\}). \\
\end{array}%
\end{equation*}

A suggestion for future work is to consider composite constructions of higher orders than $18$ and to form generator matrices that can be used to search for optimal binary self-dual codes of different lengths.


\begin{thebibliography}{99}

\bibitem{MAGMA} W. Bosma, J. Cannon and C. Playoust, \lq \lq The Magma algebra system. I. The user language", J. Symbolic Comput., vol. 24, pp. 235--265, 1997.

\bibitem{Bouyukliev1} I. Bouyukliev, V. Fack and J. Winna, \lq \lq Hadamard matrices of order 36", European Conference on Combinatorics, Graph Theory and Applications, pp. 93--98, 2005.

\bibitem{Dontcheva1} R. Dontcheva, \lq \lq New binary self-dual $[70,35,12]$ and binary $[72,36,12]$ self-dual doubly-even codes", Serdica Math. J., vol. 27, pp. 287--302, 2002.

\bibitem{DoughertyI} S.T. Dougherty, P. Gaborit, M. Harada, P. Sole, \lq \lq Type II codes
over $\mathbb{F}_2+u\mathbb{F}_2$", IEEE Trans. Inform. Theory, vol. 45, pp.
32--45, 1999.

\bibitem{DoughertyIII} S. T. Dougherty, J. Gildea and A. Korban, \lq \lq Extending an Established Isomorphism between Group Rings and a Subring of the $n \times n$ Matrices", \textbf{In press.}

\bibitem{Korban2} S.T. Dougherty, J. Gildea, A. Korban and A. Kaya, \lq \lq Composite Constructions of Self-Dual Codes from Group Rings and New Extremal Self-Dual Binary Codes of length 68", Advances in Mathematics of Communications, vol. 14, pp. 677--702, 2020. 

\bibitem{Korban3} S.T. Dougherty, J. Gildea, A. Korban and A. Kaya, \lq \lq New Extremal Self-Dual Binary Codes of length 68 via Composite Construction, $\mathbb{F}_2+u\mathbb{F}_2$ Lifts, Extensions and Neighbours", International Journal of Information and Coding Theory (IJICOT), vol. 5, no. 3/4, 2020.

\bibitem{DoughertyII} S.T. Dougherty, J. Gildea, R. Taylor and A. Tylshchak, \lq \lq
Group Rings, G-Codes and Constructions of Self-Dual and Formally Self-Dual
Codes", Des., Codes and Cryptog., Designs, vol. 86, no. 9, pp. 2115--2138, 2018.

\bibitem{DoughertyIV} S.T. Dougherty, T.A. Gulliver, M. Harada, \lq \lq Extremal binary self dual codes", IEEE Trans. Inform. Theory, vol. 43, no. 6, pp. 2036--2047, 1997.

\bibitem{Dougherty2} S.T. Dougherty, J-L. Kim and P. Sole, \lq \lq Double circulant codes from two class association schemes", Advances in Mathematics of Communications, vol. 1, no. 1, pp. 45--64, 2007.

\bibitem{Gulliver1} T.A. Gulliver, M. Harada, \lq \lq On double circulant doubly-even  self-dual $[72,36,12]$ codes and their neighbors", Austalas. J. Comb., vol. 40, pp. 137-144, 2008.

\bibitem{Yankov2} M. Gurel, N. Yankov, \lq \lq Self-dual codes with an automorphism of order 17", Mathematical Communications, vol. 21, no. 1, pp. 97--101, 2016.

\bibitem{HurleyI} T. Hurley, \lq \lq Group Rings and Rings of Matrices", Int. Jour. Pure
and Appl. Math, vol. 31, no. 3, pp. 319--335, 2006.

\bibitem{Kaya1} A. Kaya, B. Yildiz and I. Siap, \lq \lq New extremal binary self-dual codes of length 68 from quadratic residue codes over $\mathbb{F}_2+u\mathbb{F}_2+u^2\mathbb{F}_2$", Finite FIelds and Their Applications, vol. 29, pp. 160--177, 2014.

\bibitem{selfdual72} A. Korban, All known Type~I and Type~II $[72,36,12]$ binary self-dual codes, available online at \url{https://sites.google.com/view/adriankorban/binary-self-dual-codes}.

\bibitem{Korban1} A. Korban, S. Sahinkaya, D. Ustun, \lq \lq A Novel Genetic Search Scheme Based on Nature -- Inspired Evolutionary Algorithms for Self-Dual Codes", arXiv:2012.12248.

\bibitem{RainsI} E.M. Rains, \lq \lq Shadow Bounds for Self-Dual Codes", IEEE Trans.
Inf. Theory, vol. 44, pp. 134--139, 1998.

\bibitem{Yildiz1} N. Tufekci, B. Yildiz, \lq \lq On codes over $R_{k,m}$ and constructions for new binary self-dual codes", 	Mathematica Slovaca, vol. 66, no. 6, pp. 1511--1526, 2016.

\bibitem{Yankov1} N. Yankov, M.H. Lee, M. Gurel and M. Ivanova, \lq \lq Self-dual codes with an automorphism of order 11", IEEE, Trans. Inform. Theory, vol. 61, pp. 1188--1193, 2015.

\bibitem{Zhdanov1} A. Zhdanov, \lq \lq New self-dual codes of length 72", arXiv:1705.05779.

\bibitem{Zhdanov2} A. Zhdanov, \lq \lq Convolutional encoding of 60, 64, 68, 72-bit self-dual codes", arXiv:1702.05153.

\end{thebibliography}
\end{document}